A planar magneto-inductive lens for three-dimensional subwavelength imaging

M. J. Freire and R. Marqués

Departamento de Electrónica y Electromagnetismo, Facultad de Física, Universidad de Sevilla, Avda. Reina Mercedes s/n, E 41012 Seville, Spain

**Abstract**

A planar near-field magnetoinductive lens operating in the microwave range is presented. The proposed device consists of two parallel planar arrays of metallic broadside coupled (BC-) split ring resonators (SRRs), or BC-SRRs. Power coming from a point-like source located in front of the lens is focused into a receiver located in free space behind the device. This focus is clearly removed from the back side of the lens, and has a size which is an order of magnitude smaller than the free space wavelength of the incoming radiation. The imaging properties of the device mainly relies on the excitation of magnetoinductive surface waves on the BC-SRR arrays. By simply scaling the BC-SRRs size, as well as the arrays periodicity, the frequency of operation of the device can be tuned over a wide frequency range. Thus the proposed design is potentially useful for many applications ranging from megahertzs to terahertzs.

After the seminal papers of Veselago [1] and Pendry [2] on the imaging properties of left-handed slabs, there has been a great deal of interest on this subject, including the design of sub-diffraction imaging devices [2]. In fact, lenses overcoming diffraction limits could find application not only in imaging devices, but also in the recording of information, microwave heating and many other technological areas. However, it soon becames apparent that losses and dispersion – both unavoidable in any practical left-handed material – would strongly limit in practice the performances of these devices [3-5]. In particular, losses will prevent subwavelength imaging at distances higher than a small number of wavelengths in any practical lens [4,5]. Therefore, attention will soon focus in the near-field imaging of electromagnetic signals. A way to obtain quasi-perfect imaging in the quasi-electrostatic limit is by using a slab of negative dielectric permitivitty ($\varepsilon$) [2]. Since many metals show this property at optical frequencies, thin metallic slabs could be good candidates for obtaining near-field subwavelength imaging at these frequencies. Quasi-perfect near-field imaging can be also obtained by using magnetized ferrite slabs operating in the microwave range [6]. It should be emphasized that these two last proposals are only theoretical and have not been experimentally confirmed until present.

Regarding practical realizations of subdiffraction imaging devices, they are presently restricted to the microwave range [7,8], making use of planar circuit analogous of left-handed media [7] or of very thin (single-cell depth) highly anisotropic left-handed slabs in free space [8]. The first realization [7] seems not to open the way to any possible generalization leading to three-dimensional practical lenses. The second one, apart from being highly anisotropic, does not enable to reproduce images "of finite depth" [8], the imaging being restricted to planes parallel to the lens. Microwave three-dimensional imaging was already reported in [9], although the size of the focus was not of sub-wavelength dimensions. Thus, it can be concluded that devices producing three-dimensional isotropic sub-wavelength imaging have not been reported before.

In this letter we present a planar magnetoinductive lens for near-field imaging in the microwave range. As in previous works [2,6-8], the key concept of the proposed imaging process is the amplification, inside the lens, of the evanescent Fourier harmonics (FHs) coming from the source. This amplification produces the restoration of the amplitude of each FH at the image plane so that the source field is reproduced at this plane. In negative-$\varepsilon$ imaging devices [2] and in magnetized ferrite slabs [6], this amplification is due to the excitation of surface waves (surface plasmons in metals and magnetostatic surface waves in ferrites) at the shadowed interface of the slab. These surface waves, which decay from this interface towards both the source and the image, are coupled to the decaying evanescent FHs coming from the source, in such a way that the amplitude of these FHs is restored at the image plane [2,6]. This mechanism is sketched in Fig.1.a, where the decaying–growing–decaying process is illustrated for a single FH. Recently a similar mechanism of amplification of evanescent modes has been reported in a pair of coupled resonator arrays [10]. To achieve ideal perfect imaging, this process should take place for all the FH coming from the source. Since the lens operates at a single frequency, this implies that the dispersion relation for the surface waves at the shadowed interface of the lens (the plane $z=0$ in Fig.1.a), should be very flat, so that surface waves corresponding to whatever value of the wavevectors ($k_x, k_y$) can be simultaneously excited at the frequency of operation of the lens [2,6,10]. This condition is in fact fullfilled by the ideal surface plasmons supported by a

negative-$\varepsilon$ slab interface [2]. It is also fullfilled by the magnetostatic surface waves supported by a magnetized ferrite slab interface [6].

In the proposed device (see Fig.2), magnetoinductive surface waves are excited at the shadowed lens interface. Magnetoinductive waves were first reported by Shamonina et al. [11] in both one- and two-dimensional arrays of split rings resonators (SRRs). Magnetoinductive waves are produced by the inductive coupling between SRRs, due to the strong magnetic dipoles induced at each resonator near its resonance. Magnetoinductive waves behave as quasi-magnetostatic surface waves in many aspects, as it was noted in [12]. The dispersion relation for a planar square array of SRRs as those shown in Fig.2 can be written as [11,12]

$$\frac{\omega_0^2}{\omega^2} = 1 + 2\frac{M}{L}\Big[\cos\big(k_x a\big) + \cos\big(k_y a\big)\Big] \tag{1}$$

where, accordingly to [12], only the coupling between the nearest neighbours has been taken into account. In (1) $L$ and $M$ accounts for the SRRs self-inductance and mutual inductance respectively, and $a$ is the array periodicity. The fractional bandwidth $\Delta\omega/\omega_0$, predicted by (1) is

$$\frac{\Delta\omega}{\omega_0} \approx 4\frac{M}{L} \tag{2}$$

where $\Delta\omega$ is the bandwidth of excitation of the magnetoinductive surface waves. This bandwidth is very small, of the order of a 10% in the configuration shown in Fig.2. Therefore, magnetoinductive waves fullfill very approximately the aforementioned condition of having a "flat" dispersion relation, which is needed for subwavelength imaging. Since magnetoinductive waves are quasi-magnetostatic waves, the magnetic field, $\boldsymbol{H}$, can be obtained from the quasi-static magnetic potential $\Psi_m$: $\boldsymbol{H} = \boldsymbol{-grad}\,\Psi_m$.

The magnetostatic potential satisfies Laplace's equation in free space. Thus, the $z$-dependence for the magnetoinductive waves that can be excited at the SRR square arrays is of the kind $\Psi_m (k_x,k_y,z) = \Psi_{m,0} \, exp(-k\zeta)$, where $\Psi_{m,0}$ is a constant, $k=(k_x^2+k_y^2)^{1/2}$, and $\zeta$ is the distance to the SRR array. Regarding the behavior of $\Psi_m$ near the SRR array, it differs from that of the surface plasmons and ferrite magnetostatic potentials [2,6]. On the one hand, the magnetostatic potential $\Psi_m$ should experience a discontinuity at each lens interface, equal to the mean magnetic dipole moment per square meter: $\Delta\Psi_m = N<m_z>$ ($<m_z>$ is the average magnetic dipole of the SRRs and $N$ is the number of SRRs per square meter). On the other hand, the normal component of the magnetic field, $H_z$, should be continuous across these interfaces. The behavior of $\Psi_m$ for the considered magnetoinductive surface waves is sketched in Fig.1.b. Although this behavior differs from that reported in [2] and [6] for the surface plasmons and for the magnetostatic surface waves, respectively, it still can produce FHs amplification. This amplification mechanism is sketched in Fig.1.b (it is worth noting that, if the absolute values of the magnetostatic potential were plotted, this graphic will be identical to Fig.1.a). In Figs. 1.a and 1.b, it can be seen how a single FH coming from the source (at z= -3d/2) is coupled to a surface wave excited at the shadowed lens interface (at z=0), and how its amplitude is finally restored at the focus plane (z=d/2). It is worth to note that the potential behavior at both sides of z=-d reproduces that of the companion *mathematical* solution of the surface wave generated at z=0. For an isolated SRR array this solution will be *non-physical*, since the magnetostatic potential will grow to infinity at both sides of the array. However, the presence of the source at z=-3d/2, as well as that of the companion SRR array at z=0, allows for the excitation of this field configuration around the SRR array at z=-d.

Fig. 2 shows the practical device analyzed in this paper. The lens consists of two parallel square arrays of SRRs separated by a distance *d* (a layer of Rohacell[TM] foam was placed between the layers containing the arrays in order to provide mechanical stability). Broadside coupled SRRs (BC-SRRs) are used instead of conventional SRRs in order to avoid cross-polarization effects, and to reduce the electrical size of the resonators [13]. The BC-SRRs are photo-etched on a dielectric substrate of thickness *h*, as it is shown in the inset of Fig.2. Two square loop antennas are used as source and receiver, respectively. These square loops are also fabricated by photo-etching a metallic pattern on a dielectric substrate. The size of these loops is taken equal to four times the unit cell area of the BC-SRRs array, in order to average the detailed contribution to the total field coming from each individual resonator. The input antenna (the source) is fixed at a distance *d/2* from the lens; the output antenna is scanned in the *y-z* plane in order to search for the image of the input antenna. Fig.3.a shows the measured transmission coefficient between both antennas when the lens is present, and Fig.3.b shows this measurement when the lens is removed. Fig.3.a shows clearly that the transmitted power has a maximum at a distance *d/2* of the shadowed interface of the lens, i.e., at a distance *2d* of the the input antenna, in agreement with the proposed theory. Nothing similar is observed when the lens is absent. The half intensity width of the aforementioned maximum is around 10 mm in the lateral direction (y axis), and less than 2 mm in the perpendicular one (z axis). These distances are smaller than the free space wavelength at the operating frequency (about 100 mm) by an order of magnitude. The intensity of the maximun is indeed remarkable, since it is an order of magnitud higher than the intensity at the same point when the lens is removed. A noticeable fact is that the reported maximum is clearly removed from the lens shadowed interface. Thus, a three-dimensional sub-wavelength image of the source is obtained. The slight asymmetries of the field patterns shown in Fig.3 can be attributed to tolerances in the lens fabrication process, or in the positioning device used for scanning the output antenna.

In summary, subwavelength microwave three-dimensional imaging by a flat magnetoinductive microwave lens has been shown. The reported image has a sub-wavelength size, which is in agreement with the "perfect lens" theory [2], and is clearly removed from the edge of the lens. To the best of our knowledge, a similar result has not been reported before for any three-dimensional device. An advantage of the reported device over previous near-field lens designs is the simplicity of its manufacturing process, which mainly consists on etching two planar metallic patterns on a conventional circuit board, by means of standard microwave circuit manufacturing techniques. For these reasons, we feel that the reported concepts can substantially improve present approaches to electromagnetic imaging devices. Since BC-SRRs are metallic resonators with a linear response, whose frequency of resonance can be tuned by simply scaling them in space, the frequency of operation of the proposed device can be also tuned over a broad frequency range. This frequency range will be mainly limited by the usefulness of the SRR topology as a practical magnetic resonator design. Since this usefulnes seems to have been actually demonstrated until the terahertz range [14], the limits of application of the proposed design would also extend at least to these frequencies. In addition to its theoretical interest, devices based on the reported concepts can find practical applications in many technological areas, such as microwave imaging or heating, for industrial and health applications.

**Acknowledgements**

This work has been supported by DGI, Ministerio de Educacion y Ciencia (SPAIN), under project contract TEC2004-04249-C02-02.

**FIGURE CAPTIONS**

FIG. 1. a) Decaying–growing–decaying behavior of the electrostatic potential $\Psi$ in the negative-permittivity (ε) lens. b) Decaying–growing–decaying behavior of the magnetostatic potential $\Psi_m$ in the magnetoinductive (MI) lens.

FIG. 2. Experimental setup for the measurement of the reported magnetoinductive microwave lens. The lens consists of two parallel square planar dielectric substrates with an area of 7x7 cm$^2$, separated by a foam slab of thickness $d$=4mm. A periodic two-dimensional array of broadside coupled split rings resonators (BC-SRRs) is photoetched on the two interfaces of each substrate. The square inset in the figure shows in detail the geometry of the BC-SRR, which has an outer radius $r$=2mm and a ring width $w$=0.5mm. The periodicity of the array is 5mm. The dielectric substrate is alumina, which is commonly used in microwave circuits, with thickness $h$=0.254mm and dielectric permittivity $\varepsilon_r$=10. An input antenna and a receiving antenna are placed at the opposite sides of the lens. Both antennas are squared loops with an area of 1x1cm$^2$ and are fabricated by photoetching a metallic pattern on an low permittivity substrate. The input antenna is fixed at a certain distance from the left interface of the lens and the receiving antenna can be scanned along the Y-Z directions shown in the figure.

FIG. 3. Experimental results obtained with the setup shown in Fig. 2. The data are referred to the modulus of the transmission coefficient between the input antenna and the receiving antenna with the lens (a) and without it (b). The measurements were carried out by using a Vector Network Analyzer HP 8510 B. The input antenna is fixed at a distance of 2mm from the left interface of the lens (the location of the left interface corresponds to $z=-3d/2$ in Fig. 1.b) and its operating frequency is 3.23GHz. The receiving antenna is scanned along the Y-Z directions shown at the right side of the lens in Fig. 2. Note that in Fig. 3.a the image is formed at the point y=0mm, z=2.25mm ($z=d/2$ in Fig. 1.b), i.e., at a distance of 2.25 mm from the right interface of the lens. This distance agrees very approximately with the halfwidth of the lens, i.e., with the sum of the foam halfwidth (2 mm) and the microwave board thickness (0.254 mm). Therefore, the image formation obeys the proposed theory. Note that in Fig. 3.b. the modulus of the transmission coefficient in the point y=0mm, z=2.25mm in the absence of the lens is one order of magnitude lower than in the same point in the presence of the lens.

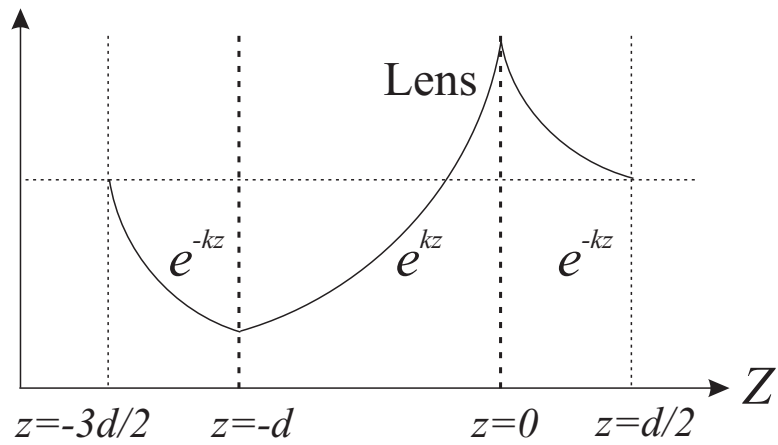

a)

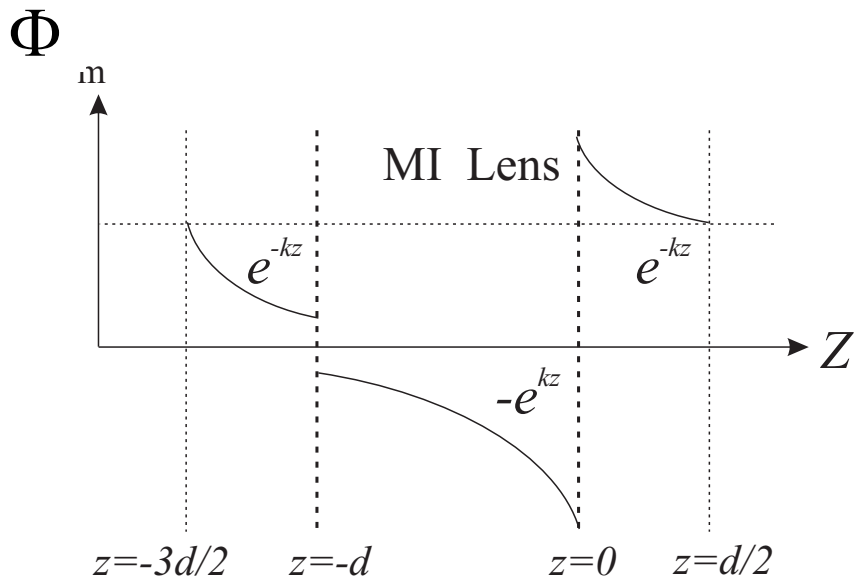

b)

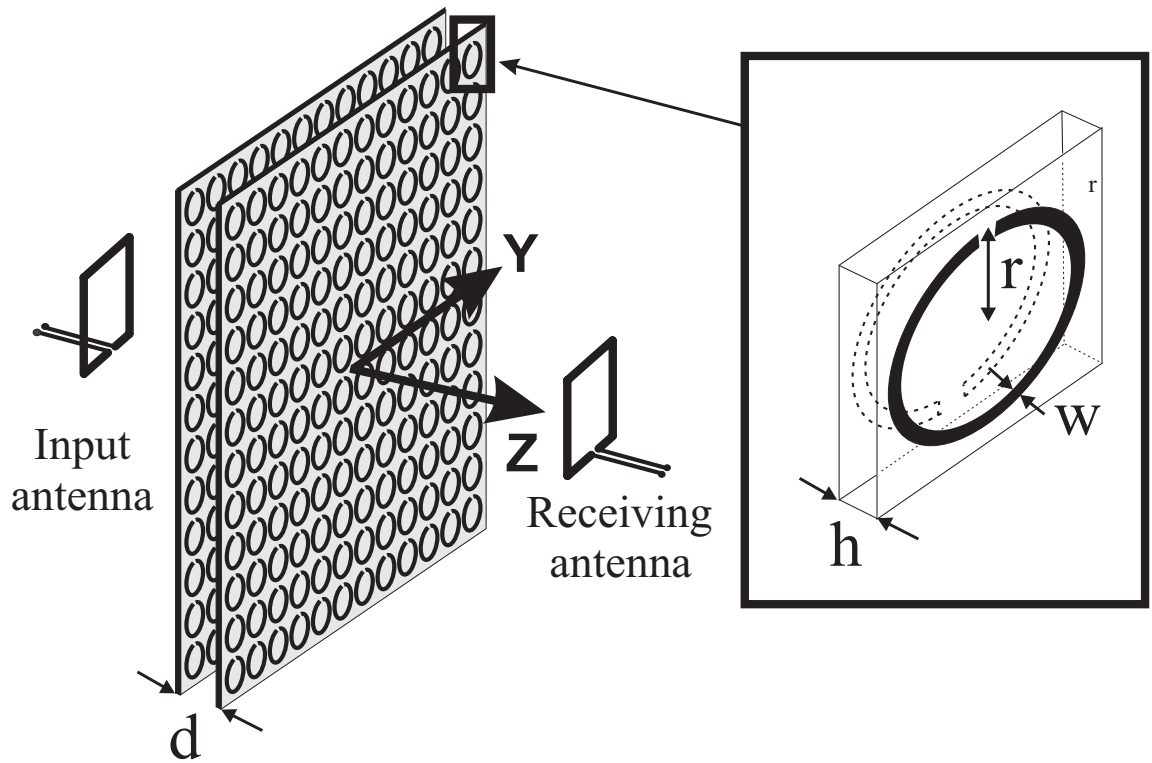

Input antenna

Receiving antenna

Figure 2, APL, M. J. Freire *et al*.

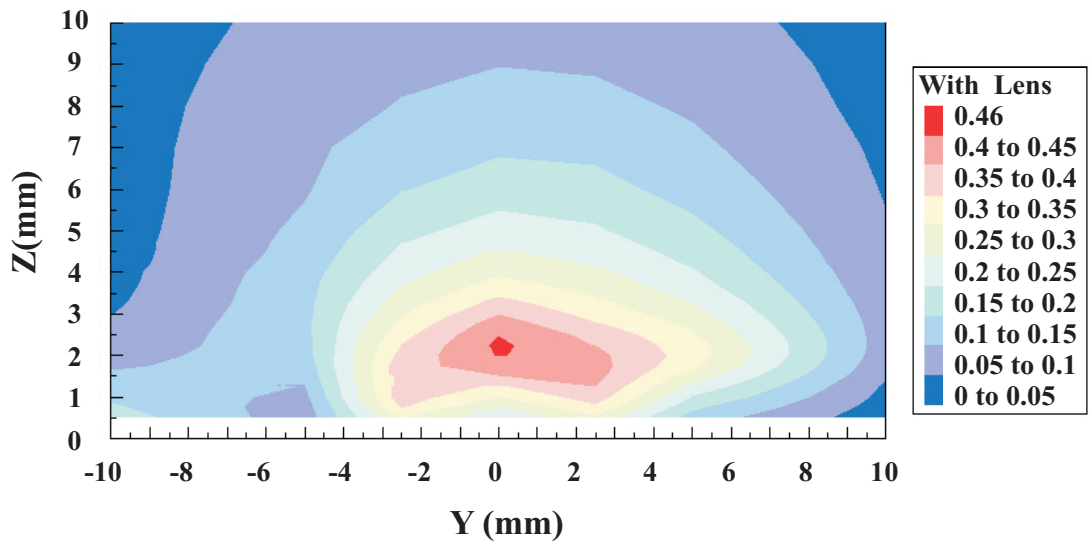

a)

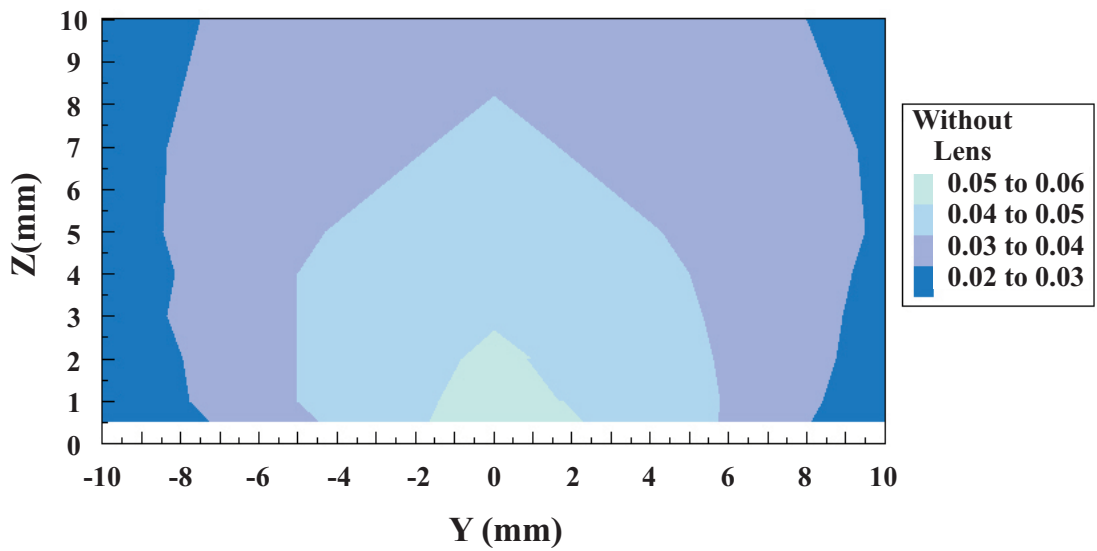

b)

**Figure 3, APL, M. J. Freire *et al*.**